\documentclass[aps,prl,twocolumn,nofootinbib]{revtex4}
\usepackage{graphicx}
\usepackage{bm}
\begin{document}

\title{Anisotropy at the end of the cosmic ray spectrum?}
\author{Luis A. Anchordoqui$^a$, Haim Goldberg$^a$, and Diego F. Torres$^b$}
\affiliation{$^a$Department of Physics, Northeastern University, Boston,
MA 02115 \\
$^b$Lawrence Livermore National Laboratory, 7000 East Ave., L-413,
Livermore, CA 94550}

\begin{abstract}

The starburst galaxies M82 and NGC253 have been proposed as the
primary sources of cosmic rays with energies above $10^{18.7}$ eV.
For energies $\agt 10^{20.3}$~eV the model predicts
strong  anisotropies. We calculate the probabilities that the
latter  can be due to chance occurrence. For the highest energy
cosmic ray events in this energy region, we find that the
observed directionality has less than  1\% probability of
occurring due to random fluctuations. Moreover, during the first 5
years of operation at Auger, the observation of even half the
predicted anisotropy has a probability of less than $10^{-5}$ to
occur by chance fluctuation. Thus, this model can be subject to
test at very small cost to the Auger priors budget and, whatever the outcome of that
test, valuable information on the Galactic magnetic field will be obtained.

\end{abstract}

\maketitle

\section{Introduction}

Soon after the microwave echo of the big bang was discovered,
Greisen, Zatsepin, and Kuzmin (GZK) noted that the relic photons
make the universe opaque to cosmic rays (CRs) of sufficiently high
energy~\cite{Greisen:1966jv}. This occurs, for instance, for
protons with energies beyond the photopion production threshold
($\Delta (1232)$ resonance). After pion production, the proton (or
perhaps, instead, a neutron) emerges with at least 50\% of the
incoming energy.  A similar phenomenon (of energy degradation)
occurs for nuclei due to processes of photodisintegration.
Therefore, the characteristic attenuation length for extremely
high energy ($10^{20}~{\rm eV} \alt E \alt 10^{20.5}~{\rm eV}$)
hadrons is less than 100~Mpc, decreasing down to 10~Mpc with
rising energy~\cite{Stanev:2000fb}. The survival probability for
extremely high energy (EHE) $\gamma$-rays (propagating on
magnetic fields $\gg10^{-11}$~G) to a distance $d$, \mbox{$P(>d)
\approx \exp[-d/6.6~{\rm Mpc}]$}, becomes less than $10^{-4}$
after traversing a distance of 50~Mpc~\cite{Elbert:1994zv}. This
implies that the GZK sphere~\cite{gzk-sphere} represents a small
fraction of the size of the universe. Consequently, if the CR
sources are universal in origin, the energy spectrum should not
extend (except at greatly reduced intensity) beyond $\sim
10^{20}$~eV, a phenomenom known as the GZK cutoff. Even though
the Haverah Park~\cite{Ave:2001hq}, Yakutsk~\cite{Efimov:rk},
Fly's Eye~\cite{Bird:wp}, and HiRes~\cite{:2002ta} data show
statistically significant evidence for such a
cutoff~\cite{Bahcall:2002wi} (more than 5$\sigma$ independent of
the sample used as a basis for extrapolation), the AGASA ground
array detected a handful of events with energies $\agt
10^{20}$~eV~\cite{Takeda:1998ps}, as opposed to about 2 expected
from the GZK cutoff. Moreover, within statistical uncertainty
(which is large above $10^{20}$~eV) the flux of CRs above
$10^{18.7}$~eV reported by the AGASA
Collaboration~\cite{Takeda:1998ps} is consistent with a $E^{-2.7}$
spectrum up to the highest observed energies, suggesting that a
single acceleration mechanism is responsible for all the events
beyond that energy, unless of course a very unlikely matching of
spectra can account for the smoothness of the CR energy
distribution.

In order to analyze the effect of energy losses on the observed
spectrum, it is convenient to introduce the accumulation factor
$f_{\rm acc}$, defined as the ratio of energy-weighted fluxes for
``low'' ($10^{18.7}$ eV -- $10^{19.5}$ eV) and EHECRs. With this in mind,
if the Earth is located in a typical environment and all CR-sources have smooth
emission spectra, the observed spectrum above $10^{18.7}$~eV should
have an offset in normalization between low and EHE given by $f_{\rm acc}$.
For CR protons and nuclei with uniform distribution of sources active over
cosmological times, the cutoff due to photopion and photodisintegration
processes relates the accumulation factor to a ratio of attenuation
lengths~\cite{Farrar:2000nw} and leads to $f_{\rm acc} \sim 100$.
The smoothness of the observed CR spectrum~\cite{Takeda:1998ps},
{\it viz.} $f_{\rm acc} \sim 1$, seems to indicate that the power of nearby
sources must be comparable to that of all other sources (redshift $z>0.5$)
added together.

The simplest explanation, i.e., nearby sources are significantly
more concentrated, does not seem to be the case. Specifically, if
one simply assumes that the distribution of CR sources follows the
distribution of normal galaxies, the local overdensity is only a
factor of two above the mean, and thus insufficient to explain the
measured flux above $10^{20}$~eV~\cite{Blanton:2000dr}.
Furthermore, the arrival direction of the super-GZK events is
consistent with an isotropic distribution of sources (even when
some level of clustering was already detected~\cite{Uchihori:1999gu}),
in sharp contrast to the anisotropic distribution of light within
100~Mpc~\cite{Waxman:1996hp}. A way to avoid the problems with finding plausible 
astrophysical explanations is to look for solutions involving physics beyond the
standard model~\cite{Bhattacharjee:1998qc}. While the invocation of such new physics 
is an intringuing idea, there are now constraints that call into question
the plausibility of some of these ideas~\cite{Protheroe:1996pd}.

Recently, it was suggested that the observed near-isotropy of arrival
directions could be due to a diffuse propagation of EHECRs~\cite{Lemoine:1999ys}.
In this work, we examine specific candidate sources for this hypothesis. These are
the starburst galaxies M82 and NGC253 which have been shown to
reproduce the main features of the observed flux~\cite{Anchordoqui:2001ss}. In particular,
we study here the critical aspect of a residual anisotropy that emerge  beyond the GZK energy limit
after deflection in Galactic and extragalactic magnetic fields. Specifically,
we estimate the probability that an apparent correlation
between the arrival directions of the highest energy events and
the  two starbursts can originate as a purely random fluctuation.
After that, we study the sensitivity of Auger Observatory to the model.

\section{Diffuse propagation of cosmic rays in a magnetized neighborhood
of the Galaxy}

A popular explanation considered recently~\cite{Lemoine:1999ys} for
an isotropic distribution of arrival directions entertains the existence of large
scale intervening magnetic fields, so that even EHECRs propagate
diffusively. Indeed, there are some measurements of diffuse radio
emission from the bridge area between Coma and Abell superclusters
that under assumptions of equipartition allows an estimate of
$0.2 - 0.6~\mu$G for the magnetic field in this
region~\cite{Kim}. Such a strong magnetic field (which is
compatible  with existing upper limits on Faraday rotation
measurements~\cite{Kronberg:1993vk}) could be possibly understood
if the bridge region lies along a filament or sheet of large scale
structures~\cite{Ryu}. In light of this, it appears plausible,
though subject to verification, to assume that our Local
Supercluster contains a large scale magnetic field (say,
$10^{-8}~{\rm G} \alt B \alt 10^{-6}~{\rm
G}$~\cite{Anchordoqui:2001bs}) which provides sufficient bending
to EHECR orbits, camouflaging the exact location of the sources.

Diffusion has two distinctive regimes. Particles that are trapped
inside magnetic subdomains (of size $\ell_{\rm Mpc} \equiv
\ell/{\rm Mpc}$) follow Kolmogorov diffusion. In such a case, the
functional dependence of energy of the difussion coefficient (for
protons) is found to be~\cite{Blasi:1998xp}
\begin{equation}
D(E) \approx 0.048\,
\left(\frac{E_{20} \, \ell^2_{\rm Mpc}}{B_{\mu{\rm G}}}\right)^{1/3}\,\,
{\rm Mpc}^2/{\rm Myr}\,,
\end{equation}
where $B_{\mu{\rm G}}$ is the magnetic field strength in units of
$\mu$G and $E_{20}$ is the particle's energy in units of
$10^{20}$~eV.  With rising energy, the Larmor radius of the
particles starts approaching $\ell$ and there is a transition to
Bohm diffusion. The diffusion coefficient in this regime is of
order the Larmor radius times velocity ($\sim c$).

If CRs propagate diffusively, the radius of the sphere for potential proton
sources becomes significantly reduced. This is because one expects negligible
contribution to the flux from times prior to the arrival time of the
diffusion front, and so
the average time delay in the low energy region,
\begin{equation}
\tau_{\rm delay} \approx \frac{d^2}{4 D(E)}\,,
\end{equation}
must be smaller than the age of the source, or else the age of the
universe (if no source within the GZK radius is active today, but
such sources have been active in the past). Note that the diffuse
propagation of EHE protons requires  magnetic fields $\sim 1\mu$G.
Therefore, for typical coherence lengths of extragalactic magnetic
fields ($\ell \sim 1$~Mpc) the time delay of CRs with $E \approx
10^{18.7}$~eV cannot exceed $\tau_{\rm delay} \alt 14$~Gyr,
yielding a radius of $d \sim 30$~Mpc. In the case CR sources
are active today, the radius for potential sources is even
smaller $d \sim 5$~Mpc.  Centaurus A, at a distance of 3.4 Mpc and
galactic coordinates $l = 310^\circ, \, b = 20^\circ$, is the
nearest active galaxy, and the only one within a distance of
5~Mpc. Phenomenological arguments identify Centaurus A as a
plausible progenitor of all CRs observed on Earth with energies
$\agt 10^{18.7}$~eV~\cite{Farrar:2000nw,Anchordoqui:2001nt}.
However, detailed numerical simulations seem to indicate that
large scale magnetic fields ${\cal O} (\mu$G) cannot provide
sufficient angular deflection to explain all the observational
data: (1) the large deflection angle of the highest energy event
recorded by the Fly's Eye experiment (see Table I) with respect to
the line of sight to Centaurus A must be explained as a $2\sigma$
fluctuation~\cite{Isola:2001ng}, (2) for an emission spectrum
$\propto E^{-2.4}$ and maximum injection energy of $10^{21}$~eV,
the angular power spectrum shows a $3\sigma$ quadrupole deviation
from AGASA observations~\cite{Isola:2002ei}.

If magnetic fields in the nanogauss range exist in the
neighborhood of the Galaxy, it is possible that ultrahigh energy
cosmic ray nuclei could diffuse sufficiently in order to attain
the observed near-isotropy. For a CR nucleus of charge $Ze$ in a
magnetic field $B_{\rm nG} \equiv B/10^{-9}$~G, the Larmor radius
is
\begin{equation}
R_L \approx \frac{100 \,\,E_{20}}{Z\,\,B_{\rm nG}}\,\,{\rm Mpc}\,\,.
\end{equation}
In this case, the sphere of potential sources is severely
constrained by the GZK cutoff: less than 1\% of iron nuclei (or
any surviving fragment of their spallations) can survive more
than $3 \times 10^{14}$~s with an energy $\agt 10^{20.5}$~eV.
Therefore, the assumption that EHECRs are heavy nuclei implies
ordered extragalactic magnetic fields $B_{\rm nG} \alt 15 - 20$,
or else nuclei would be trapped inside magnetic subdomains
suffering catastrophic spallations. There are two candidate
sources within the GZK sphere; namely, the nearby ($d \sim
3$~Mpc) metally-rich starburst galaxies M82 ($l = 141^\circ$, $b
= 41^\circ$) and NGC253 ($l =89^\circ$, $b =
-88^\circ$)~\cite{Anchordoqui:1999cu}.  Phenomenological
considerations based on analytical estimates of the diffusion
coefficient and approximations to the photodisintegration losses
and angular deflections suggest that the power of these starbursts
is enough to provide all CRs observed on Earth above
$10^{18.7}$~eV~\cite{Anchordoqui:2001ss}. This analytical study is
consistent with Monte Carlo simulations~\cite{Bertone:2002ks}.
Specifically, the spectrum observed by AGASA can be fitted with a
single source located at $d = 3.2$~Mpc if the spectrum of nuclei
is $\propto E^{-1.6}$. This is a hard spectrum compared to the
expected $E^{-2}$ from the Fermi mechanism. In this context, it
should be noted that the fit in \cite{Bertone:2002ks} has strong
statistical weight from points near 10$^{19}$ eV. However, there
is significant systematic uncertainty in the observed energy
spectrum in this region. For example, as recently 
noted~\cite{Bahcall:2002wi}, a downward shift of 11\% in the AGASA
energy calibration is required in order to bring the resulting
spectrum into agreement with Fly's Eye data. This softening of
the observed energies will require a steeper Fermi-like injection
spectrum.

\begin{table}
\caption{The highest energy cosmic rays. The energy resolution for the
AGASA experiment was taken
from Ref.~\cite{Takeda:1998ps}.}
\begin{tabular}{cccccc}
\hline\hline
Date & Experiment & \,\,$E$[EeV]\,\, & \,\,$l$\,\, & \,\,$b$\,\,& Ref.\\
\hline
89/05/07 & Yakutsk & $300^{+100}_{-178}$ & $162.0^\circ$ & $\phantom{-0}2.0^\circ$ & \cite{Efimov:pw}\\
91/10/15 & Fly's Eye & $320 \pm 90$ & $163.4^\circ$ & $\phantom{-0}9.6^\circ$ & \cite{Bird:1994uy}\\
93/12/03 & AGASA & $213 \pm 75$ & $ 130.5^\circ$ & $-41.4^\circ$ &
\cite{Hayashida:1994hb}\\
01/05/10 & AGASA & $280 \pm 98$ & $106.3^\circ$ & $-39.0^\circ$ & \cite{Sakaki}\\
\hline \hline \label{table}
\end{tabular}
\end{table}

{\it The most salient feature of the starburst hypothesis is the
prediction of an anisotropy at the high end of the spectrum}. For
an extragalactic, smooth, magnetic field of $\approx 15 -20$~nG,
diffusive propagation of particles below $10^{20}$~eV evolves to
nearly complete isotropy in the CR arrival
directions~\cite{Anchordoqui:2001ss,Bertone:2002ks}. Above this
critical energy there is a transition range (up to $10^{20.3}~{\rm
eV}$) where the combined bending in extragalactic and Galactic
magnetic fields leads to loss of  directionality. With rising
energy, the average deflection in the extragalactic magnetic field
is significantly reduced, and is roughly $10^\circ \alt \theta
\alt 20^\circ$ in the energy range $10^{20.3}~{\rm eV} \alt E \alt
10^{20.5}~{\rm eV}$~\cite{Bertone:2002ks}. In order to
incorporate typical uncertainties in energy
resolution as well
as those in the Monte Carlo simulation~\cite{Bertone:2002ks}, we
will increase the upper limit of this deflection to 30$^\circ$.
Heavy nuclei suffer additional deflection in the Galactic magnetic field.

The large scale structure of the Galactic magnetic field carries
substantial uncertainties, because the position of the solar
system does not allow global measurements. The average field
strength can be directly determined from pulsar observations of
the rotation and dispersion measures average along
the line of sight to the pulsar with a weight proportional to the
local free electron density, $\langle B_{||} \rangle \approx 2 \mu$G~\cite{Manchester}.
Measurements of polarized
synchrotron radiation as well as Faraday rotation of the radiation
emitted from pulsars and extragalactic radio sources revealed that
the global structure of the magnetic field in the disk of our
Galaxy could be well described by spiral fields with $2 \pi$
(axisymmetric, ASS) or $\pi$ (bisymmetric, BSS)
symmetry~\cite{Beck}. In the direction perpendicular to the
Galactic plane the fields are either of odd (dipole type) or even
(quadrupole type) parity. Discrimination between these models is
complicated. Field reversals are certainly observed (in the
Crux-Scutum arm at 5.5 kpc from the Galactic center, the
Carina-Sagittarius arm at 6.5 kpc, the Perseus arm at 10 kpc, and possibly another beyond~\cite{Han:1999vi}).
However, as discussed by Vall\'ee~\cite{Vallee}, turbulent dynamo
theory can explain field reversals at distances up to $\sim$~15~kpc
within the ASS configuration. Interestingly, if the
Galactic field is of the ASS type, CRs entering the Galaxy with $l
< 180^\circ$ are deflected towards increasing values of $l$ and
decreasing values of $|b|$~\cite{Stanev:1996qj}. Consequently, as
we show in what follows, each arrival direction given in Table I
can be traced backwards to one of the starbursts.

\begin{figure}
\begin{center}
\includegraphics[height=8.5cm]{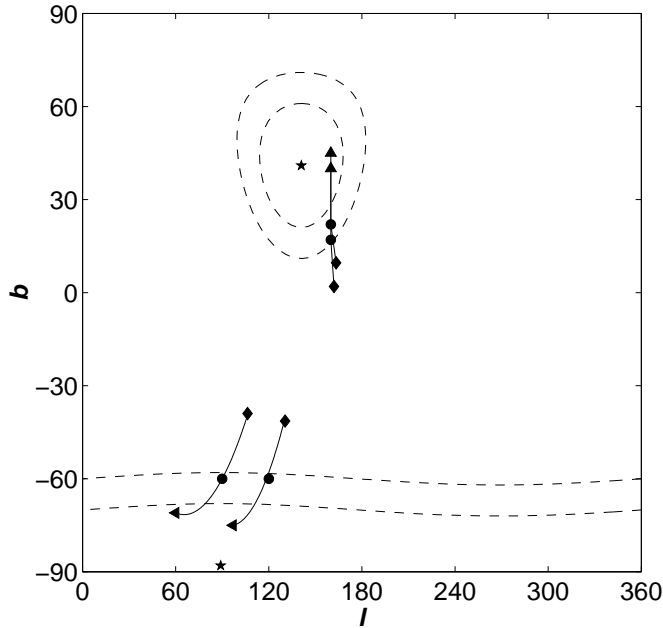}
\caption{Directions in Galactic coordinates of the four highest
energy cosmic rays at the boundary of the Galactic halo. The diamonds
represent the observed incoming directions. The circles and arrows
show the directions of neon and iron nuclei, respectively, before
deflection by the Galactic magnetic field. The solid line is the
locus of  incoming directions at the halo for other species with
intermediate atomic number. The stars denote the positions of M82
and NGC253. The dashed lines are projections in the $(l,b)$
coordinates of angular directions within $20^{\circ}$ and
$30^{\circ}$ of the starbursts.} \label{magnetica}
\end{center}
\end{figure}

The field strength in the Galactic plane ($z=0$) for the ASS
model is generally described by~\cite{Stanev:1996qj}
\begin{equation}
B (\rho, \theta) = B_0(\rho) \, \cos^2 [\,\theta - \beta \ln (\rho/\xi_0)]\,,
\end{equation}
where $\theta$ is the azimuthal coordinate around the Galactic center
(clockwise as seen from the north Galactic pole), $\rho$ is the galactocentric
radial cylindrical coordinate, and
\begin{equation}
B_0(\rho) = \frac{3 r_0}{\rho} \, {\rm tanh}^3 (\rho/\rho_1)\,\,\,\mu{\rm G}\,.
\end{equation}
Here, $\xi_0 = 10.55$~kpc stands for the galactocentric distance
of the maximum of the field in our spiral arm, $\beta = 1/\tan p$
(with the pitch angle, $p =-10^\circ$), $r_0 = 8.5$~kpc is the
Sun's distance to the Galactic center, and $\rho_1 = 2$~kpc. The
$\theta$ and $\rho$ coordinates of the field are correspondingly,
\begin{equation}
B_\theta = B(\rho, \, \theta) \,\cos p\,\,, \hspace{1cm} B_\rho = B(\rho, \theta) \, \sin p\,.
\end{equation}
The field strength above and below the Galactic plane (i.e., the dependence on $z$)  has a
contribution coming from the disk and another from the halo,
\begin{eqnarray}
B (\rho, \theta, z) & = & B(\rho, \theta) \,\,\, {\rm tanh} (z/z_3)\, \nonumber \\
 & \times &  \left( \frac{1}{2 \cosh (z/z_1)} +
\frac{1}{2 \cosh (z/z_2)}\right) \,\,,
\label{oko}
\end{eqnarray}
where $z_1 = 0.3$~kpc, $z_2 = 4$~kpc and $z_3 = 20$~pc.  Figure~\ref{magnetica}
shows the extent to which   the observed arrival
directions of the CRs listed in Table I deviate from their
incoming directions at the Galactic halo because of  bending in
the magnetic field given in Eq.~(\ref{oko}). The incoming CR
trajectories are traced backwards  up to distances of 20 kpc away
from the Galactic center, where
the effects of the magnetic field
is negligible. The diamond at the head of each solid line denotes
the observed arrival direction, and the points along these lines
indicate the direction from which different nuclear species (with
increasing mass) entered the Galactic halo. In particular, the
tip of the arrows correspond to incoming directions at the halo
for iron nuclei, whereas the circles correspond to nuclei of neon.
Additionally shown in the figure, indicated by stars, is the
location of the two starbursts.  Regions  within the dashed lines
comprise  directions lying within   $20^\circ$ and $30^\circ$
degrees of  the starbursts. It is  seen that trajectories for CR
nuclei  with  $Z\ge 10$  can be further traced back to one of the starbursts,
within the uncertainty of the extragalactic deviation.

The trajectories in Fig.~\ref{magnetica} result from motion in
the regular component of the Galactic magnetic field. However, there is some 
evidence supporting the existence of a random component
roughly comparable in magnitude to the regular component~\cite{Beck:2000dc}. Thus, 
using the random walk formulation~\cite{Waxman:1996zn} with
coherence lengths of $\sim 1$~kpc, we estimate that the
trajectories should be broadened by an angle $ \delta \vartheta \sim
40^{\circ}\ (Z/20).$ The effect of this broadening on our
analysis will be further discussed in the following section.

The effects of the BSS configuration are completely different.
Because of the averaging over the frequent field reversals,
the resulting deviations  of the CR trajectories are markedly smaller, and in
the wrong  direction for correlation of current data with the
starburst sources. We note that the energy-ordered 2D correlation
distribution of the AGASA data is in disagreement with
expectations for positively charged particles and the BSS
configuration~\cite{Alvarez-Muniz:2001ur}.

\section{Assessment of random coincidences}

We now attempt to assess to what extent these correlations are
consistent with chance coincidence. To do so, we first observe
that the angular deviation of a CR arrival direction at the outer
edge of the galaxy with respect to the straight line of sight is
roughly
\begin{equation}
\theta \approx 0.3^\circ \,\,\, \frac{d}{{\rm kpc}} \,\,\,\frac{\langle B_{||}
\rangle}{\mu{\rm G}} \,\,\,
\frac{Z}{E_{20}}\,\,.
\end{equation}
Hence, for $d \approx 5 - 10$~kpc, the average deflection of heavy
nuclei with energies in the range $10^{20.3}~{\rm eV} \alt E \alt
10^{20.5}~{\rm eV}$ is $30^\circ \alt \theta  \alt 40^\circ$.  We
arrive at the effective angular size of the source in a two-step
process. Before correcting for bias due to the coherent structure
of the Galactic magnetic field, the deflections in the
extragalactic and Galactic fields (regular and random components)
may be assumed to add in quadrature, so that the angular sizes of
the two sources are initially taken as cones with opening
half-angles between 40$^{\circ}$ and 60$^{\circ}$, which for the
purpose of our numerical estimate we  approximate to  50$^\circ.$
However, the global structure of the field will introduce a
strong bias in the cosmic ray trajectories, substantially
diminishing the effective solid angle. The combined deflections
in the $l$ and $b$ coordinates mentioned above concentrate the
effective angular size of the source to a considerably smaller
solid angle~\cite{Stanev:1996qj}. As a conservative estimate, we
retain 25\% of this cone as the effective source size. A clear
prediction of this consideration is that {\em the incoming flux
shows a strong dipole anisotropy in the harmonic decomposition.}

\begin{figure}
\begin{center}
\includegraphics[height=10.5cm]{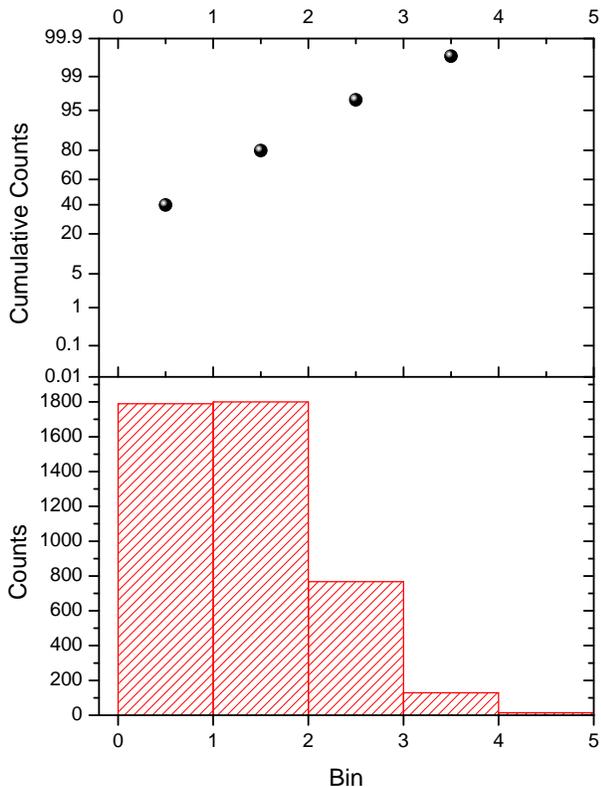}
\caption{Simulation results for 4500 trials (a larger number of
trials do not modify these results). 99\% of the simulations
yield results with less than 4 random coincidences, as described
in the text.} \label{fig}
\end{center}
\end{figure}

In order to assess the likelihood of a random occurrence of the
predicted spatial distribution we performed numerical simulations
in the spirit of Ref.~\cite{Romero:1999tk}. By randomly generating
four CR positions in the portion of the sky accessible to the
existing experiments (declination range $\delta > -10^\circ$), an
expected number of random coincidences can be obtained. The term
``coincidence'' is herein used to label a synthetic CR whose
position in the sky lies within an effective solid angle
$\Omega_{\rm eff}$ of either starburst. $\Omega_{\rm eff}$ is
characterized by a cone with opening half-angle reduced from
$50^{\circ}$ to $24^{\circ}$ to account for the 75\% reduction in
effective source size due to the magnetic biasing discussed above.
Cosmic ray positional errors were considered as circles of
$1.6^\circ$ radius for AGASA. For the other experiments the
asymmetric directional uncertainty was represented by a circle
with radius equal to the average experimental error.
Figure~\ref{fig} presents the simulation results. There are, as
we have seen, four real coincidences. However, the random
prediction for the mean number of coincidences is $0.81\pm
0.01$~\cite{sd}. The Poisson probability~\cite{Goldberg:2000zq}
for the real result to be no more than the tail of the random
distribution is just
\begin{equation}
P(\geq 4)= \sum_{k=4}^{\infty} \frac {\lambda^k \exp^{-\lambda} }
{k!} = 9\times 10^{-3}, \label{result}
\end{equation}
where $\lambda$ is the mean value of the random results. Alternatively,
we may analyze this in terms of confidence intervals. For the four observed
events, with zero background, the Poisson signal mean 99\% confidence interval is
$0.82-12.23$~\cite{Feldman:1997qc}. Thus our observed mean for random
events, $0.81 \pm 0.01$, falls at the lower edge of this
interval, yielding a 1\% probability for a chance occurrence.

We now discuss the implications of our results. Clearly, spatial
correlation analysis with a well defined and large sample of CR
positions ought to provide the key to the identification of EHECR
sources. The result embodied in Eq.~(\ref{result}) is not
compelling enough to definitively rule out chance probability as
generating the correlation of the observed events with the
candidate sources, but it is suggestive
enough to deserve serious attention in analyses of future data.
Besides, it should be stressed that the starburst hypothesis predicts a
spectrum which is approximately a smooth power law between $10^{18.7}$~eV and $10^{20.5}$~eV,
in very good agreement with that reported by the AGASA Collaboration~\cite{Anchordoqui:2001ss}.
In addition, we note that a medium mass nucleus fits the shower profile of the highest energy
Fly's Eye event quite well~\cite{Halzen:1994gy}. Moreover, the high muon density observed in
the Yakutsk event also favors a nucleus primary~\cite{Efimov:pw}.

\section{Low budget anisotropy target for the highest energy cosmic rays}

The superior angular and energy resolution of the Pierre Auger
Observatory~\cite{Anchordoqui:2002hs} will allow the high end of
the energy spectrum and the CR arrival directions to be measured
with unprecedented precision. The protocols for testing
anisotropy detection claims are being restricted to hypotheses
that must be specified {\it a priori} in order to ensure: (1) that
the sample is not (inadvertently) devised to suit a special
hypothesis after a  preliminary study of the data, and (2) that
the number of potential sources is not so large that the
criterion for negating the null hypothesis for any source is
reasonable~\cite{Clay}. Since the budgeting for candidate sources
of anisotropy is very constrained (total random probability after
accounting for all relevant trials is less than 0.001) particular
emphasis should be put -- in our view -- to those models in which
there is no new physics involved, and a plausible astrophysical
mechanism is suggested as the origin of some or all events. Some
examples are: the above mentioned Centaurus A, nearby quasar
remnants~\cite{Torres:2002bb}, and luminous infrared
galaxies~\cite{Smialkowski:ty}. In this direction, we believe the
reasons we just listed above are sufficient to encourage the
Auger community to search for evidence of the starburst model in
forthcoming measurements.

\begin{figure}
\begin{center}
\includegraphics[height=10.5cm]{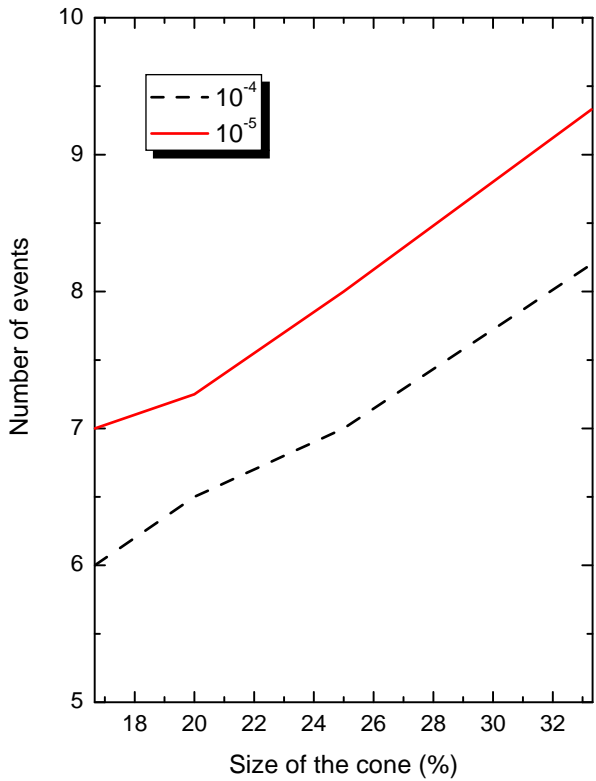}
\caption{Curves of constant probabilities in the two-dimensional
parameter space defined by the size of the cone and the minimum
number of events originating within the resulting effective solid
angle.} \label{fig2}
\end{center}
\end{figure}

We now estimate the sensitivity of Auger to our model.
The event rate for the Southern Auger Observatory
(a detector with aperture $A = 7000$~km$^2$~sr above $10^{19}$~eV, and angular resolution less
than $1.5^\circ$ with $\delta < 20^\circ$), assuming extrapolation of AGASA flux
($E^3 J(E) \approx 10^{24.5}$~eV$^2$ m$^{-2}$ s$^{-1}$ sr$^{-1}$~\cite{Takeda:1998ps}) up to
$10^{20.5}$~eV, is given by
\begin{eqnarray}
\frac{dN}{dt} & = & A\, \int_{E_1}^{E_2}\, E^3 J(E)\, \frac{dE}{E^3} \nonumber \\
  & \approx & \frac{A}{2} \,\langle E^3\, J(E) \rangle\,
  \left[ \frac{1}{E_1^2} - \frac{1}{E_2^2} \right] \nonumber \\
 & \approx & 5.3 \,\,{\rm yr}^{-1}\,\,,
\end{eqnarray}
where $E_1 = 10^{20.3}$~eV and $E_2 = 10^{20.5}$~eV. We now
consider a 5-year sample of 25 events, and note that for the
energy range under consideration the aperture of Auger is mostly
receptive to cosmic rays from NGC253. We allow for different
possibilities of the effective reduction of the cone size because
of the Galactic magnetic field biasing discussed previously. In
Fig.~\ref{fig2} we plot contours of constant probabilities ($P= 10^{-4},\
10^{-5}$) in the two-dimensional parameter space of the size of
the cone (as a fraction of the full $50^{\circ}$ circle) and the
minimum number of events originating within the resulting
effective solid angle. Several important conclusions may be
drawn. First, there is very little sensitivity of the results to
the size of the cone, the variation is less than 20\% for a 50\%
reduction in the cone size. Secondly, the model predicts that
after 5 years of operation, {\em all} of the 25 highest energy
events would be observed in the aperture described above. From
Fig.~\ref{fig2} we can see that even if 7 or 8 are observed, this is
sufficient to rule out a random fluctuation at the $10^{-5}$
level. Thus, the disproof of the starburst hypothesis can be
achieved at a very small cost, $< 10^{-5}$ out of a total
$10^{-3}$ to the Auger probability budget. Current preliminary
assignments for other hypotheses are on the order of
$10^{-4}$~\cite{Clay}.

\section{Conclusion}

We have made a definite prediction for future
observations at the Auger Observatory: if the origin of CRs above
$10^{18.7}$~eV are nearby starburst galaxies, {\em the incoming
CR flux will show a strong dipole anisotropy in the harmonic
decomposition at energies beyond $10^{20.3}$~eV.} Because of its
well-defined prediction, the model can be tested at the 5$\sigma$
level in five years of running at Auger. Therefore, we strongly
recommend that the Auger Collaboration take into account the
next-door galaxy NGC253 in their first anisotropy prescription
for super-GZK CRs. The confirmation of the starburst hypothesis
would provide, as spinoff, direct evidence for the global
structure of the Galactic magnetic field.

\hfill

\section*{Acknowledgments}
The work of L.A.A. and H.G. has been partially supported by the US
National Science Foundation (NSF), under grants No.\ PHY--0140407
and No.\ PHY--0073034, respectively. The work of D.F.T. was performed under
the auspices of the U.S. Department of Energy by University of
California Lawrence Livermore National Laboratory under contract
No. W-7405-Eng-48.



\begin{thebibliography}{99}
\bibitem{gzk-sphere} The sphere within which a source has to lie in order to
provide us with EHECRs if the primaries are subject to the GZK
cutoff phenomenon.
\bibitem{sd} The standard deviation of the 4500 trial sample is 0.8, giving a standard deviation
of 0.01 for the mean.
\end{thebibliography}
\end{document}